\documentclass[namedreferences]{solarphysics}
\usepackage[optionalrh,solaromanenum]{spr-sola-addons} % For Solar Physics 
\usepackage{graphicx}                    % For eps figures, newer & more powerful
\usepackage{rotating}
\usepackage{color}                       % For color text: \color command
\usepackage[urlcolor=blue,breaklinks=true]{hyperref}
\usepackage{breakurl}                         % For breaking URLs easily trough lines
\usepackage{footmisc}
\usepackage{threeparttable}	% For including footnotes in tables
\usepackage{adjustbox}
\usepackage{subfig}

\newcommand{\solphys}{{\it Solar Phys.}}

\newcommand{\apj}{    {\it Astrophys. J.}}
\newcommand{\apjs}{    {\it Astrophys. J. Suppl.}}
\newcommand{\apjl}{    {\it Astrophys. J. Lett.}}
\newcommand{\aap}{    {\it Astron. Astrophys.}}

\newcommand{\ssr}{    {\it Space Sci. Rev.}}

\newcommand{\procspie}{    {\it Proc. SPIE}}
\newcommand{\lya}{Ly$\alpha$}

\newcommand{\expectedsuccessrate}{e}
\newcommand{\measuredsuccessrate}{m}
\newcommand{\enf}{n_\textrm{expected}}
\newcommand{\mnf}{n_\textrm{measured}}
\begin{document}
\begin{article}

\begin{opening}

\title{On the Performance of Multi-Instrument Solar Flare Observations During Solar Cycle 24}

\author{Ryan~O.~\surname{Milligan}$^{1,2,3,4}$, Jack~\surname{Ireland}$^{3,5}$}

\runningauthor{R.O. Milligan, J. Ireland}
\runningtitle{On the Performance of Multi-Instrument Solar Flare Observations During Solar Cycle 24}

\institute{$^{1}$ School of Physics and Astronomy, University of Glasgow, Glasgow, G12 8QQ, UK \\
			$^{2}$ Astrophysics Research Centre, School of Mathematics and Physics, Queen's University Belfast, University Road, Belfast, BT7 1NN, Northern Ireland \\
            $^{3}$ Solar Physics Laboratory, Heliophysics Division, NASA Goddard Space Flight Center, Greenbelt, MD 20771, USA \\
            $^{4}$ Department of Physics Catholic University of America, 620 Michigan Avenue, Northeast, Washington, DC 20064, USA \\
            $^{5}$ ADNET Systems, Inc., 6720B Rockledge Drive, Suite 504,
Bethesda, MD 20817, USA. \\
              email: \url{ryan.milligan@glasgow.ac.uk} \\ 
              }

\begin{abstract}
The current fleet of space-based solar observatories offers us a wealth of opportunities to study solar flares over a range of wavelengths. Significant advances in our understanding of flare physics often come from coordinated observations between multiple instruments. Consequently, considerable efforts have been, and continue to be made to coordinate observations among instruments (\textit{e.g.} through the \textit{Max Millennium Program of Solar Flare Research}). However, there has been no study to date that quantifies how many flares have been observed by combinations of various instruments. Here we describe a technique that retrospectively searches archival databases for flares jointly observed by the \textit{Ramaty High Energy Solar Spectroscopic Imager} (RHESSI), \textit{Solar Dynamics Observatory} (SDO)/\textit{EUV Variability Experiment} (EVE) (\textit{Multiple EUV Grating Spectrograph} (MEGS)-A and MEGS-B), \textit{Hinode}/(EUV Imaging Spectrometer, \textit{Solar Optical Telescope}, and \textit{X-Ray Telescope}), and \textit{Interface Region Imaging Spectrograph} (IRIS). Out of the 6953 flares of GOES magnitude C1 or greater that we consider over the 6.5 years after the launch of SDO, 40 have been observed by six or more instruments simultaneously. Using each instrument's individual rate of success in observing flares, we show that the numbers of flares co-observed by three or more instruments are higher than the number expected under the assumption that the instruments operated independently of one another. In particular, the number of flares observed by larger numbers of instruments is much higher than expected. Our study illustrates that these missions often acted in cooperation, or at least had aligned goals. We also provide details on an interactive widget (\textsf{Solar Flare Finder}) now available in \textsf{SSWIDL} that allows a user to search for flaring events that have been observed by a chosen set of instruments. This provides access to a broader range of events in order to answer specific science questions. The difficulty in scheduling coordinated observations for solar-flare research is discussed with respect to instruments projected to begin operations during Solar Cycle 25, such as the \textit{Daniel K. Inouye Solar Telescope}, \textit{Solar Orbiter}, and \textit{Parker Solar Probe}.
\end{abstract}


\end{opening}

\section{Introduction} 
\label{s:intro}
The study of solar flares is a high-priority research area in the international heliophysics community. Understanding the physics of these energetic events is crucial, not only for the field of space weather, but also in the broader scope of astrophysics where similar processes are believed to occur in stellar flares, black-hole accretion disks, and in the Earth's magnetotail. Observations of solar flares are made by many different instruments, both in space and on the ground. These instruments provide imaging, photometric, and spectroscopic data over a range of wavelengths, from radio waves through the optical and EUV to X-rays and $\gamma$-rays: often the greatest advances in our understanding of solar flares come through various combinations of these datasets.
From \citet[Section 7.2]{flet11}:
\begin{quote}
The multifarious observations across the broad spectrum of phenomena each help us to characterize the equilibrium change in the corona and chromosphere that we call a flare, and it should be clear that the multiwavelength approach is crucial in flare studies. It tells us where the flare energy starts and where it ends up, and something about the intermediate steps. It also provides some geometrical and diagnostic information about the flare magnetic environment, at different levels in the atmosphere, and how and when this changes as the flare proceeds. This big picture cannot be reached using one spectral region on its own. The multiwavelength observations have many detailed applications as we try to understand specific mechanisms that are at work in various phases and regions of the flare development.
\end{quote}

However, it is difficult to keep track of which flares have been observed by which instruments. While most currently operational missions have their own individual flare lists (\textit{e.g.} Hinode Flare Catalog: \citealt{wata12} or \textit{Solar Dynamics Observatory} (SDO)/\textit{EUV Variability Experiment} (EVE): \citealt{hock12}), it was only recently that the first inter-instrument catalog became available, hosted by New Jersey Institute of Technology (NJIT)\footnote{\url{solarflare.njit.edu}} \citep{sady17}. The \textit{Max Millennium Program for Solar Flare Research} (see \citealt{bloo16} for a recent review) and others have aimed to coordinate ground- and space-based instrumentation to observe a flaring active region simultaneously in order to optimize the scientific return. However, this can be difficult due to factors such as coordinating across multiple time zones, planning schedules being uploaded days in advance, ground-based seeing conditions, competing scientific priorities, and so on.

Therefore when a solar flare is known to have been observed by a combination of instruments, the event can receive considerable attention as a consequence. A notable recent example of this is the 29 March 2014 X-class flare, which was observed by four space-based observatories and one ground based telescope\footnote{\url{www.nasa.gov/content/goddard/nasa-telescopes-coordinate-best-ever-flare-observations/}}.
Consequently there have been 23 refereed publications that discuss this flare, according to a NASA ADS fulltext search on the Solar Object Locator keyword \textsf{SOL2014-03-29}. Similarly, the first X-class flare of Solar Cycle 24 (\textsf{SOL2011-02-15}) was simultaneously observed by multiple instruments at high cadence, resulting in 42 refereed publications to date. The exceptional data coverage of each of these events allowed \cite{klei16} and \cite{mill14}, respectively, to investigate the redistribution of nonthermal electron energy. They were both able to compare radiative losses in the chromosphere across a range of wavelengths with the energy injected by nonthermal particles from hard X-ray observations. In both cases only 15\,--\,20\,\% of the nonthermal energy could be accounted for from longer-wavelength measurements. Understanding where this ``missing energy' went to can only be answered by even better data coverage. Clearly there is great scientific merit in multi-instrument observations of the same event.

Likewise for other astronomical research areas where coordinated observations of transient objects (``Targets Of Opportunity'') at different wavelengths are highly desirable. The study of supernovae, for example, is facilitated by the availability of both lightcurves (to understand the evolution) and spectra (to understand the composition and dynamics). The \textit{Open Supernova Catalog} \citep{guil17} acts as a central repository providing access to data for over 42,000 known supernova events. According to the statistics page of the website\footnote{\url{sne.space/statistics/}} only 6\,\% of these events have \textit{both} photometric and spectroscopic data available (34\,\% only have lightcurves, 7\,\% only have spectra, and 53\,\% have neither). 

This article presents an analysis of flare statistics by retrospectively cross-referencing metadata from a suite of instruments that take flare-relevant observations -- the \textit{Ramaty High Energy Solar Spectroscopic Imager} (RHESSI: \citealt{lin02}), the \textit{Multiple EUV Grating Spectrograph} (MEGS; \citealt{crot04}) -A and -B components of EVE \citep{wood12}, the \textit{EUV Imaging Spectrometer} (EIS: \citealt{culh07}), the \textit{Solar Optical Telescope} (SOT: \citealt{tsun08}), the \textit{X-Ray Telescope} (XRT: \citealt{golu07}), and the \textit{Interface Region Imaging Spectrometer} (IRIS: \citealt{depo14}) -- to search for flaring events observed simultaneously, either intentionally or serendipitously. The purpose of this article is to present an overview of how successful the solar community has been in capturing flare data through coordinated efforts. We also describe a database of these events that give researchers access to multi-wavelength datasets with which to address a given science question. Section~\ref{s:data_anal} describes how the various archives from each instrument were exploited. Section~\ref{s:results} presents the findings. The conclusions and a discussion are presented in Sections~\ref{s:conc} and \ref{s:disc}, respectively.
%, while Section~\ref{appendixa} describes an interactive widget accessible through SSWIDL

\section{Data Analysis}
\label{s:data_anal}

In order to cross-reference datasets from different instruments to infer which  observed a given solar flare simultaneously, it is important to define what exactly constitutes a flare. The most commonly accepted catalog is that of the \textit{Geostationary Operational Environmental Satellite} (GOES) event list provided by NOAA/SWPC. This defines a solar flare as a continuous increase in the one-minute averaged X-ray flux in the long-wavelength channel (1\,--\,8\,\AA) of the GOES \textit{X-ray Sensor} (XRS: \citealt{hans96}) for the first four minutes of the event. The flux in the fourth minute must be at least 1.4 times the initial flux. The start time of the event is then defined as the first of these four minutes. The peak time is when the long-wavelength channel flux reaches a maximum, thus defining its class. The end of an event is defined as the time when the long channel flux reaches a level halfway between the peak and initial values\footnote{\url{www.swpc.noaa.gov/products/goes-x-ray-flux}}. However in the vast majority of instances the NOAA catalog does not provide information on the location of a flare on the solar disk. As this is necessary for cross-referencing with the pointing information for reduced field-of-view instruments, the location of each flare was determined from the SSW Latest Events list, which is accessible through the Heliophysics Events Knowledgebase\footnote{\url{www.lmsal.com/hek/}} (HEK). Flare locations are determined by subtracting the SDO/\textit{Atmospheric Imaging Assembly} (AIA: \citealt{leme12}) 131\,\AA\ image closest to the GOES start time, from that image closest to the GOES peak time. The flare location is then extracted from the peak intensity of this difference image (S. Freeland; private communication, 2017). Knowing the timing and position of each event then allowed this information to be cross-referenced with the metadata from other instruments to determine whether or not they observed the same location at the same time. \textit{Note that this does not guarantee that a given instrument actually detected flaring emission, but only that the timing and pointing of a given dataset were consistent with the timing and location of the flare.} B-class flares were not included in this study due to discrepancies between flare locations derived from RHESSI and SDO/AIA and were therefore deemed unreliable. The SSW Latest Events list also has several months of data missing\footnote{October\,--\,December 2012; July\,--\,November 2013; May 2014; February 2015; March and June 2016.}. Nevertheless, out of the 8090 flares  of GOES class C1 or greater that appear in the NOAA event list, 6953 (86\,\%) are also in the SSW Latest Events list and include location information.

\begin{figure}[!t]
\begin{center}
\includegraphics[width=\textwidth]{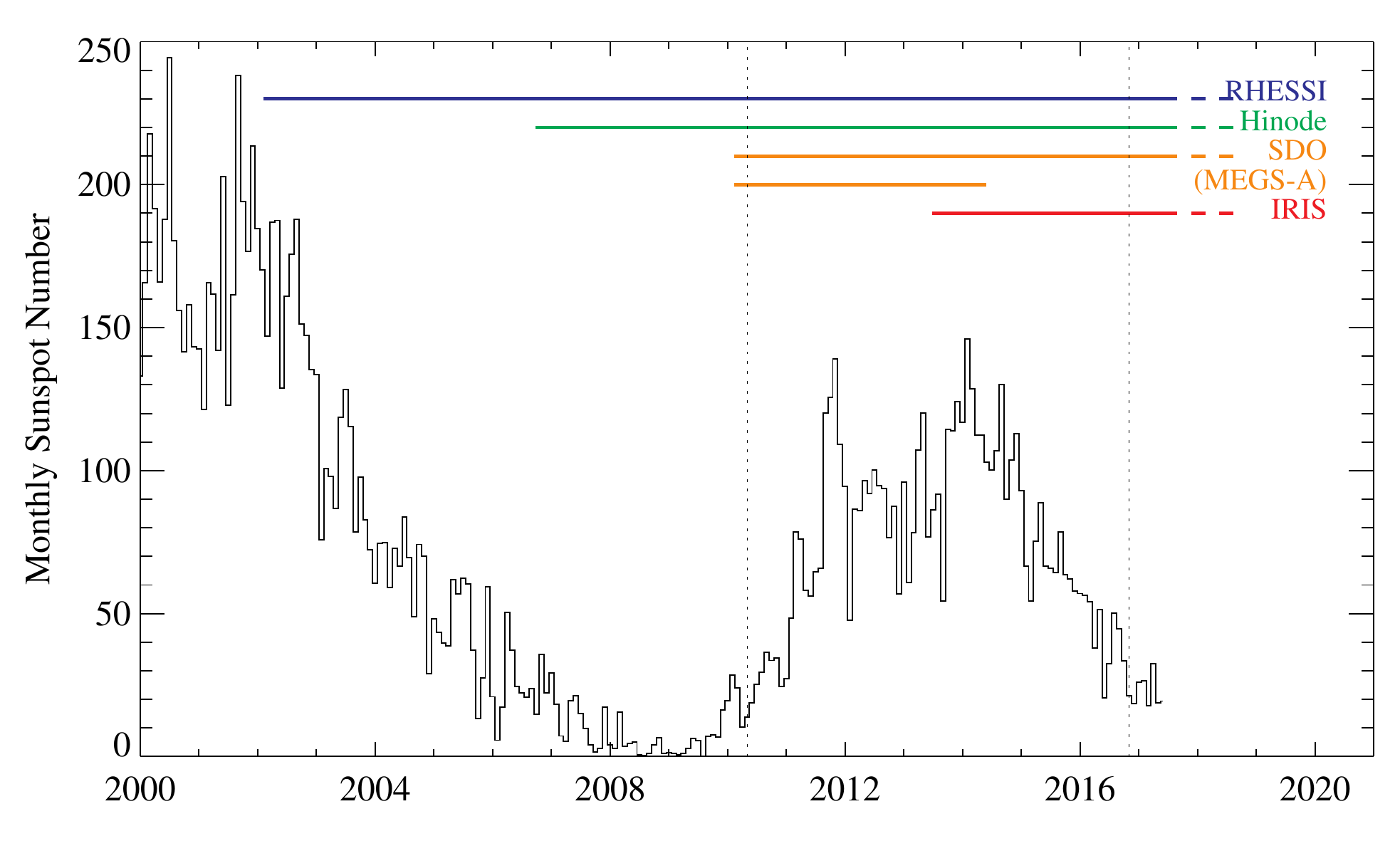}
\caption{Solar Cycles 23 and 24 (average monthly sunspot number) with mission durations overplotted. The two vertical-dotted lines denote the 6.5-year time range considered for this study. Note that SDO/EVE MEGS-A and IRIS only overlapped for $\approx$11 months.}
\label{solar_cycle}
\end{center}
\end{figure}

For the purposes of this study, only flares greater than GOES C1 class that occurred over the 6.5 years of Solar Cycle 24 observed by SDO \citep{pesn12} were considered. This defines the date range 1 May 2010 to 31 October 2016, as denoted by the vertical-dotted lines in Figure~\ref{solar_cycle}. Also shown are the durations of the missions considered in this study. Note that EVE MEGS-A and IRIS were only operational together for around 11 months after IRIS was launched, and before MEGS-A suffered a power anomaly on 26 May 2014\footnote{\url{lasp.colorado.edu/home/eve/2014/05/28/eve-megs-a-power-anomaly/}}. 

\begin{figure}
\begin{center}
%\begin{sideways}
%\includegraphics[width=\textwidth]{20140329_142600_C3_hsi100_megsab_eis_sot_xrt_iris.png}
\includegraphics[width=\textwidth]{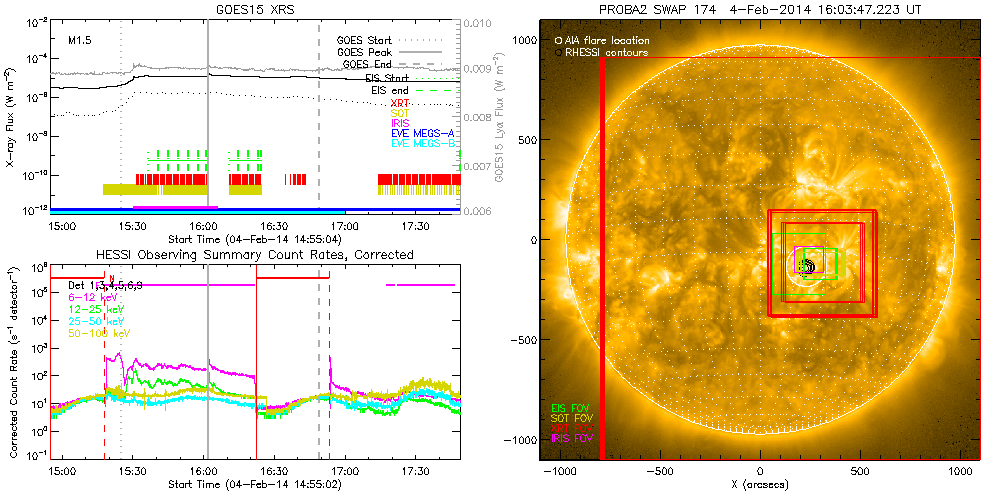}
\caption{Sample event from this study that was observed by all instruments; an M1.5 flare that occurred on 4 February 2014. Upper left panel: GOES/XRS lightcurves in 1\,--\,8\,\AA\ (solid black curve) and 0.5\,--\,4\,\AA\ (dotted-black curve), along with the GOES/EUVS-E (\lya) profile in gray. Vertical dotted, solid, and dashed grey lines denote the start, peak, and end times of the GOES event, respectively. Dotted- and dashed-green ticks mark the start and end times of each \textit{Hinode}/EIS raster, respectively, while red and yellow ticks mark the times of each SOT and XRT image, respectively. Horizontal blue and cyan lines illustrate the times which MEGS-A and MEGS-B were exposed, respectively, while the horizontal purple line shows the time of the corresponding IRIS study. Lower-left panel: RHESSI lightcurves up to the maximum energy detected, with GOES start, peak, and end times overlaid. Right panel: A PROBA2/SWAP 174\,\AA\ image taken near the peak of the flare. The white circle is 100$''$ wide centered on the location derived from AIA 131\AA\ images, while the black contours mark out the 6\,--\,25\,keV emission observed by RHESSI. The fields of view of EIS, SOT, XRT, and IRIS are overplotted in green, yellow, red, and purple, respectively.}
\label{sff_plot}
%\end{sideways}
\end{center}
\end{figure}

Figure~\ref{sff_plot} shows a sample plot from the \textsf{Solar Flare Finder} widget, which was developed in tandem with this study (see Appendix~\ref{appendixa}). Plots such as this have been generated for every SSW event since the launch of SDO, and they are being continuously updated. These plots allow the user to readily view the timing and pointing of each instrument during a chosen event. This particular plot shows one flare from this study that was found to have been observed by all seven instruments: an M1.5 flare that occurred on 4 February 2014. The upper-left panel shows the GOES X-ray lightcurves with the start, peak, and end times overlaid (vertical gray-dotted, solid, and dashed lines, respectively). Note that for completeness, the time profiles of GOES EUVS-E (centered on the Lyman-$\alpha$ -- \lya\ -- line of hydrogen at 1216\,\AA; \citealt{vier07}) are also shown in gray. \cite{mill16} recently showed that these data are more reliable for flare studies than the EVE MEGS-P data given that the GOES/EUVS-E data exhibit a more impulsive profile -- as one would expect for chromospheric emission -- whereas current EVE MEGS-P data erroneously show a more gradually varying behavior.

\subsection{The \textit{Ramaty High Energy Solar Spectroscopic Imager}}
\label{ss:rhessi}
RHESSI, launched on 5 February 2002\footnote{The first solar flare observation was of a GOES C2 flare on 12 February 2002.}, observes the full disk of the Sun in X-rays and $\gamma$-rays. It orbits the Earth at an inclination angle of 38$^{\circ}$, at an altitude of $\approx$600~km, and as such suffers from eclipse passes and transits through the South Atlantic Anomaly. In order to determine whether or not RHESSI observed a given GOES flare event, the IDL routine \textsf{hsi\_whichflare.pro} was run between the start and end times of each flare. This searches the RHESSI flare catalog\footnote{\url{hesperia.gsfc.nasa.gov/hessidata/dbase/hessi_flare_list.txt}} for the largest event detected in the time range of interest. If a RHESSI flare is detected, the fraction of the rise time (GOES start $\rightarrow$ GOES peak) that the RHESSI flare flag was active was also calculated. While RHESSI is a full-disk instrument, its orbit implies that it may have captured anywhere from a few seconds of a given flare up to around an hour (note that some long-duration flares are detectable over several RHESSI orbits).

From the lightcurves presented in the lower-left panel of Figure~\ref{sff_plot} it can be seen that RHESSI captured the peak of the M1.5 flare up to an energy of 50\,--\,100\,keV. The contours of the RHESSI quicklook image (6\,--\,25\,keV; black contours overlaid on the EUV image) agree with the flare location computed from the AIA 131\,\AA\ data (white circle).

\subsection{The \textit{EUV Variability Experiment}}
\label{ss:sdo_eve}
The SDO spacecraft is in a geosynchronous orbit allowing it to observe the full disk of the Sun continuously without interruption (except for the occasional lunar and terrestrial eclipses). For simplicity, it was assumed that both AIA and the \textit{Helioseismic and Magnetic Imager} (HMI: \citealt{sche12}) were observing continuously throughout each event. The EVE instrument, however, is less straightforward. While MEGS-A, which provides spatially integrated Sun-as-a-star spectra over the 60\,--\,370\,\AA\ range every ten seconds, and  was exposed to the Sun continuously from launch until it ceased operations on 26 May 2014, the MEGS-B (370\,--\,1050\,\AA) and MEGS-P (\lya) exposure times have been much more erratic due to unforeseen degradation soon after launch. For much of the mission MEGS-B has only been exposed for three hours per day in order to limit degradation, as well as five minutes per hour for the consistency of long term variability studies. During periods of substantial solar activity it would observe continuously for 24\,--\,48 hours. Recently the flight software was changed to allow MEGS-B to respond to a flare trigger based on the EVE EUV Solar Photometer (ESP) flux for events $>$M1. Although there is an EVE flare catalog online\footnote{\url{lasp.colorado.edu/eve/data_access/evewebdata/interactive/eve_flare_catalog.html}}, this includes events for which MEGS-B may have only been exposed for five minutes. Therefore for the purposes of this study, MEGS-B was considered to have observed a flare if it was exposed to the Sun continuously between the GOES start and GOES peak times as determined from the daily exposure times\footnote{\url{lasp.colorado.edu/eve/data_access/evewebdata/interactive/megsb_daily_exposure_hours.html}}. However, this does not necessarily mean that the flare itself will show up in the data, as EVE is often only sensitive to flares $\gtrsim$C5 level. The times at which MEGS-A and MEGS-B were exposed to the Sun around the time of a given flare are illustrated by the horizontal blue and cyan lines, respectively, as shown in the top-left panel of Figure~\ref{sff_plot}.

\subsection{\textit{Hinode}}
\label{ss:hinode}
The \textit{Hinode} spacecraft \citep{kosu07} was launched into a Sun-synchronous orbit on 22 September 2006 and comprises three instruments: EIS, SOT, and XRT. They were designed to study the interplay between the photosphere and the corona by working in unison. However, by January 2008 \textit{Hinode} had lost the use of its X-band transmitter, resulting in a dramatic reduction in the amount of data being transmitted to the ground.

\subsubsection{The Extreme-ultraviolet Imaging Spectrometer}
\label{sss:eis}
EIS is a two-channel, normal-incidence EUV spectrometer. Its two channels cover the wavelength ranges 170\,--\,210\,\AA\ and 250\,--\,290\,\AA, selected to cover coronal emission lines with formation temperatures ranging from 8000~K (He {\sc ii}) to 16~MK (Fe {\sc xxiv}). It has a mirror that is tiltable in the solar X-direction, and is used to build up rastered spectral images of portions of the Sun in up to 25 spectral ranges. Additionally, EIS has both narrow (1$''$ and 2$''$ wide) slits, and wider (40$''$ and 266$''$ wide) imaging slots, with up to 512$''$ in the solar Y-direction. 

For this study, a flare successfully observed by EIS must have had at least one raster begin, end, or straddle the GOES start and end times as determined from the \textsf{eis\_list\_rasters.pro} routine. If such a raster exists, then all rasters within -30 minutes and +60 minutes of the GOES start and end times, respectively, are returned. The flare location as projected from AIA must have also lain within the EIS field of view. This does not imply that EIS captured any flaring emission; due to the rastering nature of the instrument, the slit may not have been over the flare site at the opportune time. In the example shown in Figure~\ref{sff_plot}, EIS was running a sequence of $\approx$three-minute rasters (denoted by the vertical green-dotted and dashed ticks) around the peak of the M1.5 flare. The associated regions of the Sun corresponding to each raster are also overlaid on the EUV image as green boxes.
%The number of flares reported in Section~\ref{s:results} is therefore an upper limit. 

\subsubsection{The Solar Optical Telescope}
\label{sss:sot}
SOT is the first large optical telescope flown in space to observe the Sun. It images sub-full-disk portions of the Sun. Its aperture is 50~cm in diameter, the angular resolution is 0.25$''$ (corresponding to 175 km on the Sun), and the wavelengths covered extend from 4800 to 6500\,\AA. SOT also includes the \textit{Focal Plane Package}, which consists of a vector magnetograph and a spectrograph. The vector magnetograph provides time series of photospheric vector magnetograms, Doppler velocity and photospheric intensity. 

In order to determine whether SOT observed a given flare, the \textsf{sot\_cat.pro} routine was run between the GOES start and end times. If the routine returned at least one image, and the flare location fell within the SOT field of view, then all corresponding images between -30 and +60 minutes of the GOES start and end times, respectively, were returned and plotted over the GOES X-ray lightcurves as shown in Figure~\ref{sff_plot} (vertical yellow ticks). The associated regions of the Sun corresponding to each SOT image are also overlaid on the EUV image as yellow boxes.

\subsubsection{The X-Ray Telescope}
\label{sss:xrt}
XRT is a high-resolution (1$''$) grazing-incidence Wolter telescope that obtains high-resolution soft X-ray images covering the energy range 0.2 to 2 keV. This reveals magnetic-field configurations and their evolution, allowing the observation of energy buildup, storage, and release process in the corona for any transient event. XRT covers a wide temperature range from 0.5 to 10 million Kelvin allowing it to see coronal features that are not visible with a normal incidence telescope. XRT can observe the full disk of the Sun, but can also return sub-full-disk images, depending on the science goal of the observation.

In order to determine whether XRT observed a given flare, the \textsf{xrt\_cat.pro} routine was run between the GOES start and end times. If the routine returned at least one image, and the flare location fell within the XRT field of view, then all corresponding images between -30 and +60 minutes of the GOES start and end times, respectively, were returned and plotted over the GOES X-ray lightcurves as shown in Figure~\ref{sff_plot} (vertical red ticks). The associated regions of the Sun corresponding to each XRT image are also overlaid on the EUV image as red boxes.

\subsection{The \textit{Interface Region Imaging Spectrometer}}
\label{ss:iris}
Launched on 27 June 2013 into a Sun-synchronous polar orbit, IRIS obtains UV spectra and images with high spatial (1/3$''$) and temporal resolution (one-second) focused on the solar chromosphere and transition region. The instrument comprises an ultraviolet telescope combined with an imaging spectrograph. IRIS records observations of material at specific temperatures, ranging from 5000~K and 65,000~K, and up to 10 MK during solar flares. IRIS is a sub-full-disk instrument, imaging portions of the solar disk and limb.

The timing and pointing of IRIS observation studies that were run during the start and end times of a given GOES event were obtained using the \textsf{iris\_obs2hcr.pro} routine. This searches the Heliophysics Coverage Registry for the \textsf{OBSID}\footnote{See \url{iris.lmsal.com/itn26/quickstart.html} for more details.} corresponding to the time of the flare, as shown by the horizontal purple line in the upper-left panel for Figure~\ref{sff_plot}. Similar to the previously mentioned instruments with limited fields of view, the pointing information obtained from the \textsf{OBSID} was cross-referenced with the flare location to determine if IRIS was pointed at the required location (purple box overlaid on the EUV image in Figure~\ref{sff_plot}).
%IRIS has about 50 basic observing modes, which are encoded in a unique identifier called \textsf{OBSID}

\begin{center}
\begin{table}[!b]
   %\captionsetup{width=\textwidth}
   \begin{adjustbox}{max width=\textwidth}
   \caption{Distribution of how many solar flares -- and of which class -- were observed by individual instruments between 1 May 2010 and 31 October 2016 based on the timing and pointing information available (where applicable). The percentage of SSW Latest Events found is calculated relative to the number of NOAA/GOES events. Percentage of flares captured by each instrument during their respective missions are calculated against the total number of events found via SSW Latest Events.
   %Note that these are upper limits and do not guarantee that flaring emission would be visible in the actual data.
   }   
   \begin{threeparttable}
   \begin{tabular}{lccccc}
   \hline
Instrument/			&			&			&			&			&Success Rate			\\
Database			&C-class	&M-class	&X-class	&Total		&Over 6.5 Years\tnote{a,b}\\
   \hline
NOAA/GOES			&7360		&685		&45			&8090		&100\,\%				\\
SSW Latest Events	&6339		&581		&33			&6953		&86\,\%					\\
\hline
RHESSI				&3673		&370		&23			&4066		&58\,\%					\\
SDO/EVE MEGS-A\tnote{a}			&3825		&343		&19			&4187	&100\%			\\
SDO/EVE MEGS-B		&787		&97			&8			&892		&12\,\%					\\
\textit{Hinode}/EIS			&496		&54			&6			&556		&8\,\%					\\
\textit{Hinode}/SOT			&1167		&177		&15			&1359		&20\,\%					\\ 
\textit{Hinode}/XRT			&3739		&357		&26			&4122		&59\,\%					\\
IRIS\tnote{b}		&523 (3349)&76 (335)   &5 (16)		&604 (3700)	&16\,\%					\\
   \hline
   \end{tabular}
   \begin{tablenotes}
   \item[a]{MEGS-A was assumed to have observed all flares from launch until it ceased operations on 26 May 2014}
   \item[b]{The total number of flares listed in the HEK between the launch of IRIS and 31 October 2016 are given in parentheses}
   \end{tablenotes}
   \end{threeparttable}
   \label{tab:instr_flares}
   \end{adjustbox}
\end{table} 
\end{center}

\section{Results}
\label{s:results}

\begin{center}
\begin{table}[!b]
   %\captionsetup{width=\textwidth}
   \begin{adjustbox}{max width=\textwidth}
   \caption{Number and percentage of total flares observed by different combinations of instruments. Note that there were 6953 flare events that were potentially observable by six or fewer instruments. Only 934 events were potentially observable by all seven of the instruments considered in this study.}
   \begin{threeparttable}
   \begin{tabular}{lcc}
   \hline
Degree							&Number of flares observed	&\% of potentially observable flares	\\
   \hline
No instruments					&127			&1.8\,\%		\\   
Exactly 1 instrument			&1432			&20.6\,\%		\\
Any 2 instruments				&2371			&34.1\,\%		\\
Any 3 instruments				&2035			&29.2\,\%		\\
Any 4 instruments				&720			&10.3\,\%		\\
Any 5 instruments				&228			&3.3\,\%		\\
Any 6 instruments				&37				&0.5\,\%		\\ 
All 7 instruments\tnote{c}		&3				&0.3\,\%		\\
   \hline
   \end{tabular}
   \begin{tablenotes}
   \item[c]{A total of 934 flares were recorded during the 11 months when both MEGS-A and IRIS were operational together.}
   \end{tablenotes}
   \end{threeparttable}
   \label{tab:joint_flares}
   \end{adjustbox}
\end{table} 
\end{center}

Based on the search criteria defined in Section~\ref{s:data_anal}, the number of flares, and their percentages of the total number of SSW Latest Events (which itself is a subset -- 86\,\% -- of the available NOAA/GOES events) that were considered to have been observed by each of the instruments are listed in Table~\ref{tab:instr_flares}. The instruments with full-disk capability and high duty cycles (RHESSI, MEGS-A, and \textit{Hinode}/XRT) unsurprisingly were able to capture more than half of the total flares considered. The remaining instruments -- which have either limited duty cycles and/or limited fields of view -- were only able to capture around 20\,\% or less of all flares during Solar Cycle 24. Similarly, the number of flares and their percentages that were observed by different combinations of instruments are listed in Table~\ref{tab:joint_flares}. Around 84\,\% of all flares were observed by between one and three instruments. Most of the remaining 16\,\% were observed by either four or five instruments, while a total of 37 flares were observed by different combinations of six instruments and only three out of 934 were observed by all seven instruments during the 11 months that they were simultaneously operating. Interestingly, 127 flares (1.8\,\%) were not observed at all by \textit{any} of the seven instruments considered.

\begin{figure}[!t]
\begin{center}
%\begin{sideways}
\includegraphics[width=\textwidth]{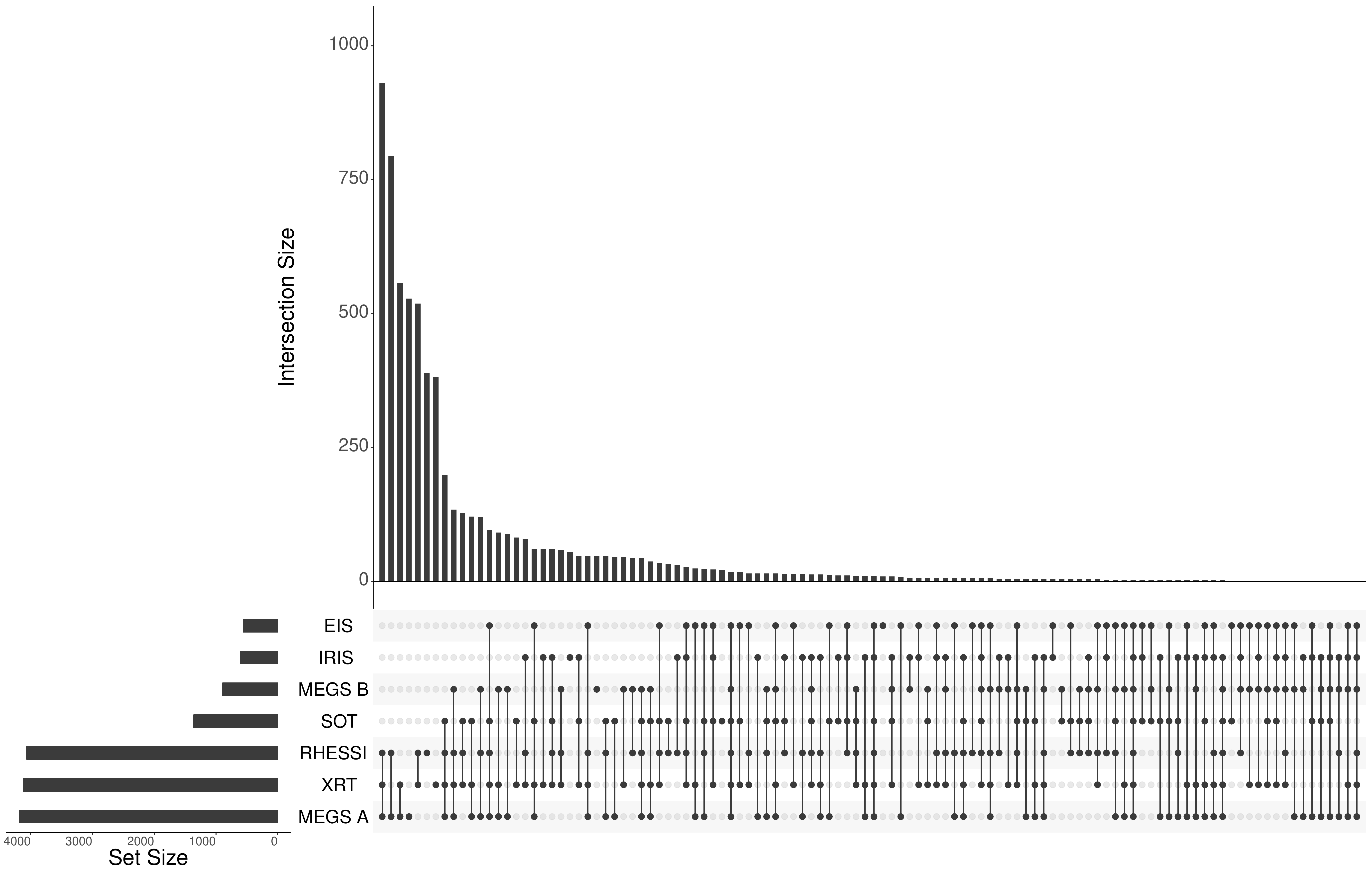}
\includegraphics[width=\textwidth]{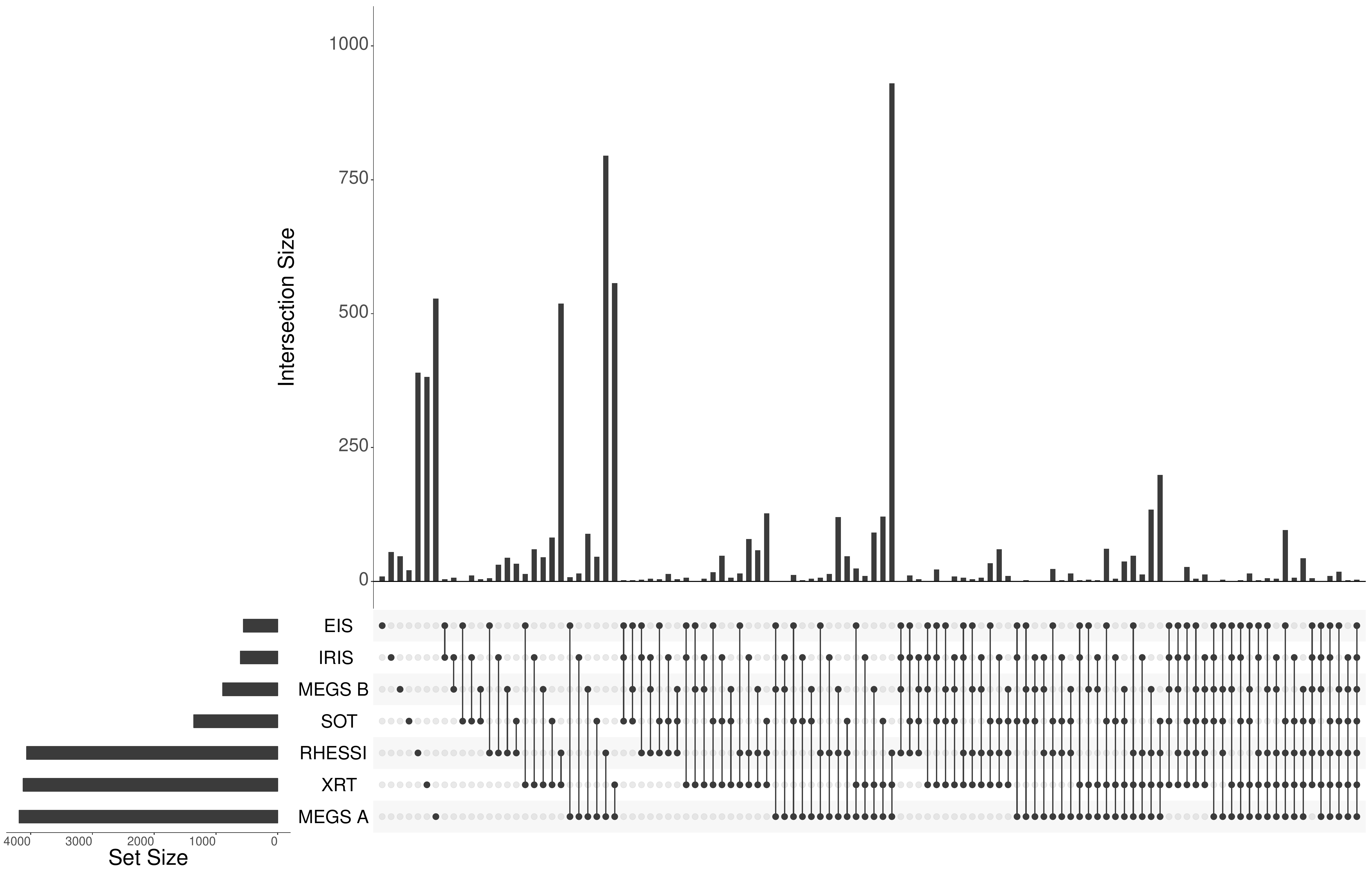}
\caption{\textsf{UpSet R} plots of the intersection of flare datasets from each instrument as ordered by decreasing frequency (top panel) and increasing number of instruments (bottom panel). Zero-element sets are not included in either plot.}
\label{f:upset}
%\end{sideways}
\end{center}
\end{figure}

The findings of how many solar flares were observed by different combinations of instruments are displayed in Figures~\ref{f:upset} as \textsf{UpSet R}\footnote{\url{gehlenborglab.shinyapps.io/upsetr/}} plots \citep{lex14}. This type of plot enables the efficient visualization of common elements of a large number of sets (the more common and familiar Venn diagram approach produces ineffective visualizations for more than $\approx$five sets). The top panel of Figure~\ref{f:upset} shows the intersections of the various combinations of datasets ordered by decreasing frequency (i.e. the most common combinations are on the left and decrease towards the right), while the bottom panel shows the same information only now ordered by increasing number of instruments (i.e. flares observed by individual instruments alone come first, with flares observed by all seven on the far right). In each plot, the total number of flares observed by each instrument are given by the horizontal black bars in the bottom-left corner. The dots connected by lines at the bottom of each figure denote the combinations of instruments considered, while the histograms above give the number of events corresponding to a given combination. The most common combination of flare datasets was RHESSI+MEGS-A+\textit{Hinode}/XRT (930 flares), due to their large fields of view and high duty cycles as mentioned above. 
\\

\subsection{Evaluation of Measured Versus Expected Success Rates}
It is difficult to give a good estimate of how many flares one would \textit{expect} to see with each instrument, given their different science goals and operational constraints. Table~\ref{tab:estimated_flares} summarizes an attempt to estimate this expectation value [$\expectedsuccessrate$] for each instrument considered in this article. The estimates are based on the average field-of-views times the duty cycles of each instrument. The area of consideration is estimated in two ways. The first estimate is simply the area of the full disk of the Sun. The second estimate assumes that there are four active regions on the Sun each with an area of $240''\times240''$, and that the majority of the duty cycle is spent examining the active-region areas. These two estimates give an upper and lower range to the percentage field of view. The percentage field of view is calculated as the percentage of the area of consideration covered by the average field-of-view of the instrument disk. The duty cycle is estimated as the percentage of the time that the instrument could have observed a flare. Crucially, the estimates assume that a \textit{random} location within the area of consideration (either the full disk of the Sun, or an estimated average area of active regions that the instrument could point to, assuming that active regions form the majority of target areas during the duty cycle). 

\begin{center}
\begin{table}[!t]
   %\captionsetup{width=\textwidth}
   \begin{adjustbox}{max width=\textwidth}
   \caption{Estimates of the percentage of flares expected to be observed [$\expectedsuccessrate$] by each instrument based on the product of their duty cycles and field-of-view. The calculation assumes that each instrument points randomly in the area of consideration. The percentage of flares that were actually observed [$\measuredsuccessrate$] is also presented.}
   \begin{threeparttable}
   \begin{tabular}{lccccc}
   \hline
Instrument					&				&			&``Expected''	&Measured	\\
							&Duty cycle		&\%FOV		&Success Rate	&Success Rate\tnote{f}	\\
                            &               &           &$\expectedsuccessrate$ &$\measuredsuccessrate$ \\
   \hline
RHESSI						&50\,\%			&100\,\%		&50\,\%			&58\,\%		\\   
SDO/EVE MEGS-A				&100\,\%		&100\,\%		&100\,\%		&100\,\%	\\
SDO/EVE MEGS-B\tnote{a}		&12.5\,\%		&100\,\%		&12.5\,\%		&12\,\%		\\
\textit{Hinode}/EIS\tnote{b}&25\,\%			&2\,--\,25\,\%	&0.5\,--\,6\%	&6\,\%		\\
\textit{Hinode}/SOT\tnote{c}&50\,\%			&1\,--\,17\,\%	&0.5\,--\,8\%	&13\,\%		\\
\textit{Hinode}/XRT\tnote{d}&100\,\%		&25\,--\,100\,\%&25\,--\,100\%	&57\,\%		\\
IRIS\tnote{e}				&100\,\%		&0.5\,--\,3\,\% &0.5\,--\,3\%	&11\%		\\ 
   \hline
   \end{tabular}
   \begin{tablenotes}
   \item[a]{Duty cycle estimated at approximately three hours per day (see text).}
   \item[b]{Duty cycle estimated by examining recent EIS planning notes. Field of view estimated at $240''\times 240''$, one quarter the full FOV of EIS.}
   \item[c]{Field of view estimated at $200''\times200''$, one quarter the full FOV of SOT.}
   \item[d]{Field of view estimated at $1024''\times1024''$, one quarter the full FOV of XRT.}
   \item[e]{Field of view estimated at $85''\times85''$, one quarter the full FOV of IRIS.}
   \item[f]{Out of the 934 flares listed in the HEK over the 11-month period that all seven instruments were operational. This is around half of all the flares listed in the NOAA/GOES event list (1774).}
   \end{tablenotes}
   \end{threeparttable}
   \label{tab:estimated_flares}
   \end{adjustbox}
\end{table} 
\end{center}

%\vspace{-0.8cm} %Not sure why this is needed. Keeps generating extra carriage returns.
A very crude estimate of the ``expected'' success rate [$\expectedsuccessrate$] is therefore the product of the duty cycle and the FOV. This can be readily compared to the measured success rate [$\measuredsuccessrate$] for each instrument individually from the 11-month period during which all seven instruments were operational together. This time period also happened to coincide with the peak of the solar cycle as illustrated in Figure~\ref{solar_cycle}. These expected and measured values are presented in the last two columns of Table~\ref{tab:estimated_flares}. The success rates over this 11-month period bear a reasonable agreement with the values measured over the entire 6.5 years under study that are presented in the final column of Table~\ref{tab:instr_flares}, and they can therefore be considered as characteristic of each instrument. They are also consistent with or better than the individual expected value implying that each pointing instrument is performing well. This reflects the fact that solar flares are a high priority science goal for these instruments, and that operators of course do not point their instruments randomly. 

The measured success rates of each individual instrument in Table~\ref{tab:estimated_flares} can be used to predict the number of flares expected to be seen by different combinations of instruments as follows. The measured success rate of each instrument indicates the probability [$\measuredsuccessrate$] of an instrument observing a flare. Therefore $1-\measuredsuccessrate$ indicates the probability of an instrument missing a flare. This suggests the use of the binomial distribution to model the number of flares detected by each instrument individually. The binomial distribution is
\begin{equation}
B(k; N, \measuredsuccessrate) = {{N}\choose{k}}\measuredsuccessrate^{k}(1-\measuredsuccessrate)^{N-k}
\label{eqn:binomial}
\end{equation}
where $k$ is the number of successful outcomes, $N$ is the number of trials and
\begin{equation}
{{N}\choose{k}}=\frac{N!}{(N-k)!k!}
\label{eqn:choose}
\end{equation}
is the binomial coefficient. Consider the case of two instruments observing the same flare. If the flare is observed by the first instrument with a probability $\measuredsuccessrate_{1}$, and by the second instrument with a probability $\measuredsuccessrate_{2}$, conditional on the first instrument having observed the flare, then the resulting probability of observing $k$ flares from a possible $N$ is 
\begin{equation}
B(k; N, \measuredsuccessrate_{1}\measuredsuccessrate_{2}).
\label{eqn:binomialcombined}
\end{equation}
This can be used to calculate $\enf$, the expected numbers of flares observed by arbitrary combinations of instruments. We define $\enf$ to be the mean value of the probability mass functions (PMF) defined by Equation \ref{eqn:binomialcombined}. This value can be compared to the actual number of flares observed [$\mnf$]. If the actual number of flares observed [$\mnf$] is much larger than the expected number due to chance [$\enf$] then this can be taken as evidence that instruments were acting in cooperation, or at least had aligned goals. The number of distinct subsets of combinations of $r$ instruments from the seven instruments that we are considering is ${{7}\choose{r}} ($Equation~\ref{eqn:choose}). For example, there are 21 possible combinations of two out of seven instruments, 35 combinations of three, and so on.

\begin{figure}[!t]
\begin{center}
\includegraphics[width=\textwidth]{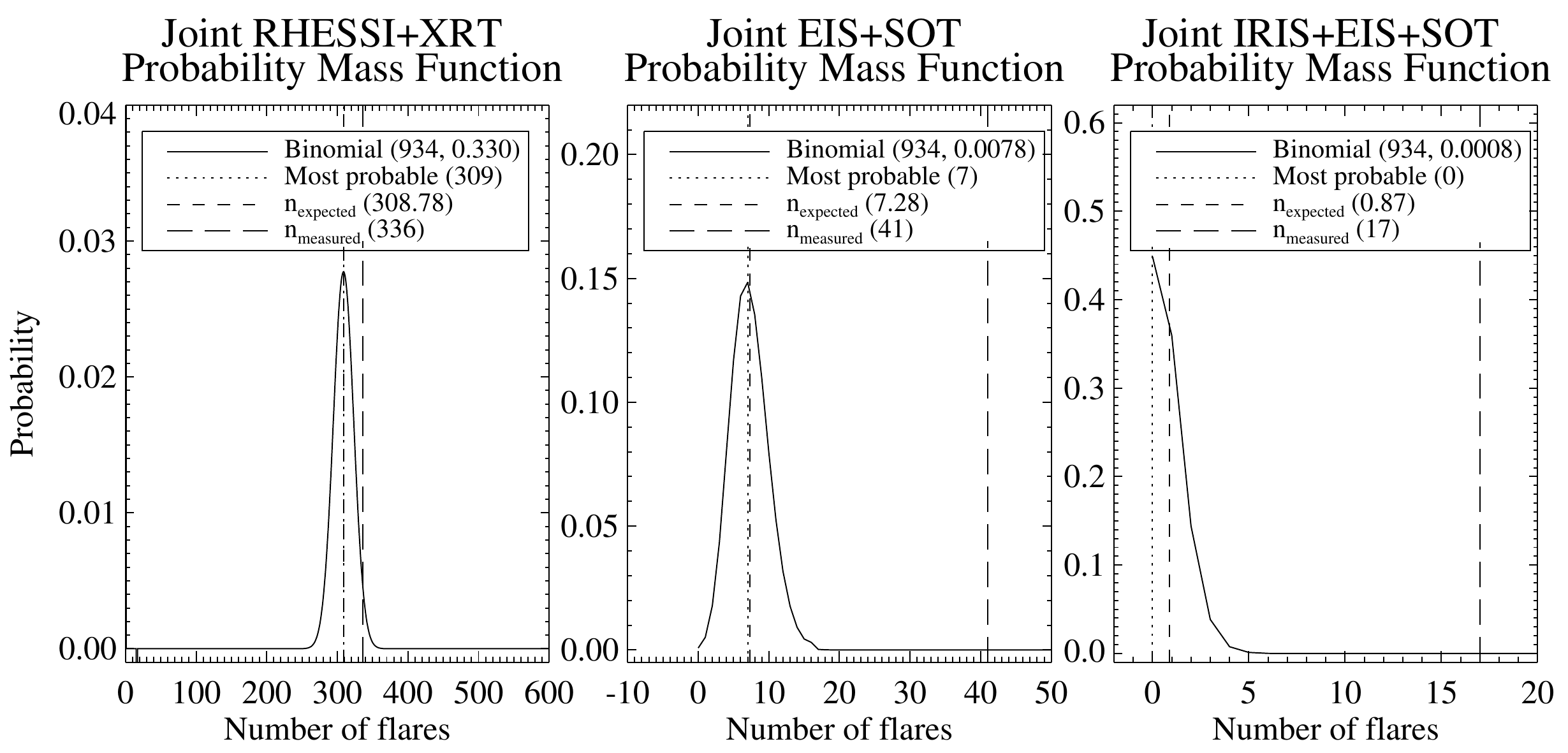}
\caption{The probability mass functions for three different combinations of instruments: RHESSI+XRT (left), EIS+SOT (center), and IRIS+EIS+SOT (right).}
\label{binomial_pmf}
\end{center}
\end{figure}

Figure~\ref{binomial_pmf} shows example PMFs for three combinations of instruments using Equation~\ref{eqn:binomialcombined}: RHESSI+XRT, EIS+SOT, and IRIS+EIS+SOT. The first two panels show that their respective probabilities display a distribution peaked on the most probable number of flares observed (309 and 7, respectively, out of 934), while the right-hand panel shows that the most probable number of flares observed simultaneously by IRIS+EIS+SOT is zero. Defining $\enf$ to be the mean of the PMF ensures that $\enf>0$. The uncertainty on $\enf$ is estimated by calculating the standard deviation of the PMF.

\begin{figure}[!t]
\begin{center}
\includegraphics[width=0.8\textwidth]{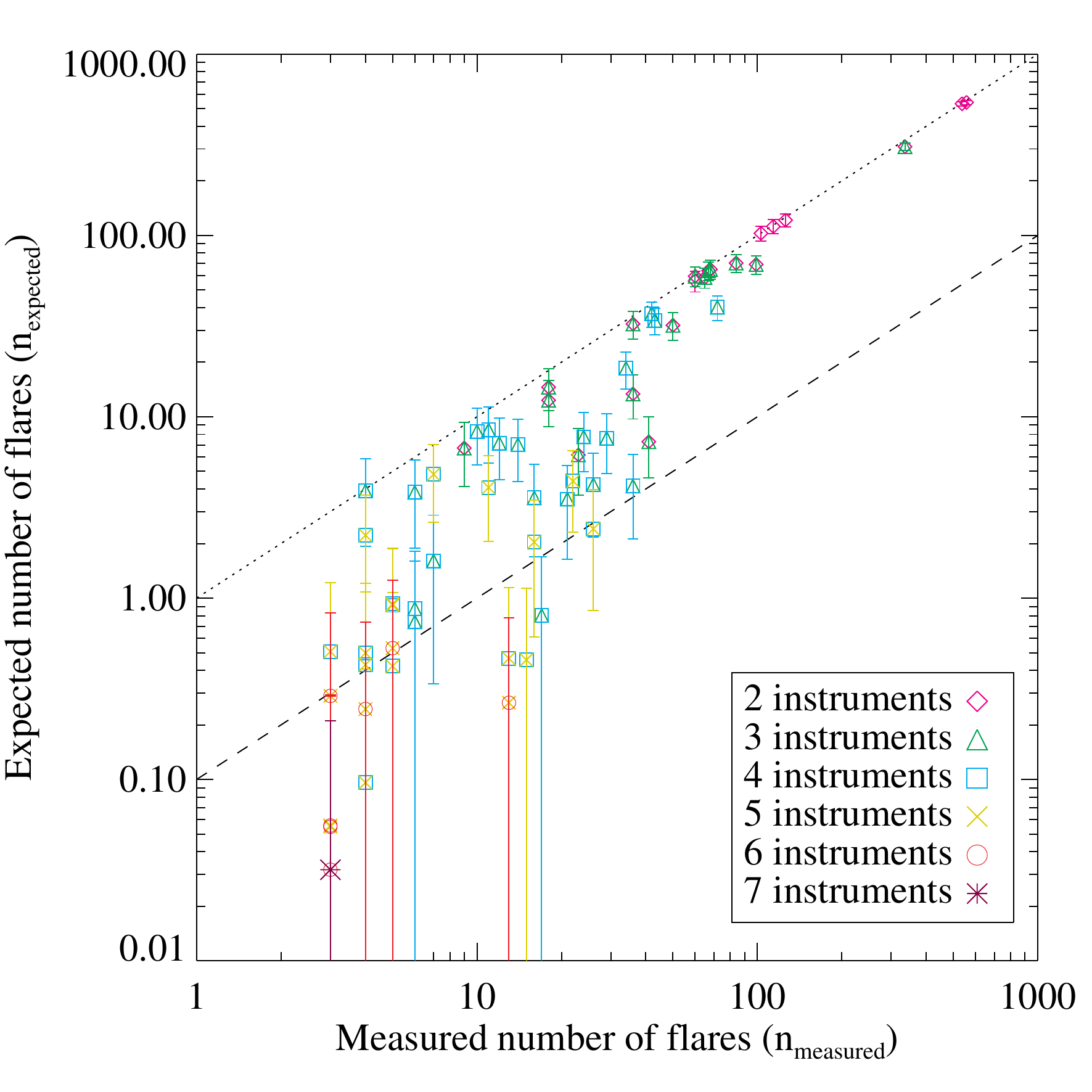}
\caption{The expected mean number of flares observed [$\enf$] by various combinations of instruments versus the number actually observed [$\mnf$]. The diagonal dotted line marks the 1:1 ratio, while the diagonal dashed line denotes the 10:1 ratio. Some of the data points overlap since some combinations of instruments are subsets of combinations of more instruments. In the case when the mean number minus the standard deviation is less than zero, the error bar is extended to the lower value of the plot range.}
\label{meas_pred}
\end{center}
\end{figure}
A scatter plot showing the expected mean number of flares out of a possible 934 that were observed [$\enf$] by 2\,--\,7 instruments, against the number actually observed [$\mnf$] is shown in Figure~\ref{meas_pred}. Points close to the 1:1 (dotted) line indicate that those combinations of instruments are co-observing flares at a rate consistent with Equation~\ref{eqn:binomialcombined}, i.e. the measured co-observation rate is close to that expected by chance.

For combinations of four or more instruments, the number of flares actually observed is far larger than that expected randomly (Equation~\ref{eqn:binomialcombined}). While the expected mean number of flares observed is small (and sometimes less than unity) for increasing numbers of coordinating instruments the number of flares actually observed is often up to ten times greater than expected if the instruments operated without co-ordination. This shows that when a flare-productive active region is present on the Sun, many instrument planners will choose to track the region -- within their operational constraints -- thereby greatly increasing the likelihood of jointly observing a given flare in conjunction with other missions. The statistics for the expected and measured number of flares observed by each possible combination of 2\,--\,6 instruments is given in Appendix~\ref{appendixb}.

\section{Conclusions}
\label{s:conc}
A statistical analysis of how many solar flares ($\geq$C1) were observed by various combinations of instruments during the 6.5 years after the launch of SDO in Solar Cycle 24 is presented. On average, over the entire 6.5 years, each flare was observed by 2.4 instruments. Out of the 6953 flares considered, only three were observed simultaneously by RHESSI, MEGS-A+B, \textit{Hinode}/EIS+SOT+XRT and IRIS: a C2.3 flare on 1 February 2014, a C4.6 flare on 3 February 2014, and an M1.5 flare on 4 February 2014. The occurrence of these three events coincided with a \textit{Max Millennium Major Flare Watch} campaign that ran from 30 January 2014\,--\,8 February 2014 on NOAA active region 11967\footnote{\url{solar.physics.montana.edu/hypermail/mmmotd/index.html}}. This illustrates the value of tracking the target region as advised by the \textit{Max Millennium Chief Observers} when trying to optimize the scientific return on solar flare datasets. Note that all seven instruments were observing contemporaneously for only 11 months, and in this time 934 events are currently listed as SSW Latest Events; therefore 0.3\,\% of all possible GOES flares were observed with all seven instruments. While this may not seem impressive, the probability of all seven instruments observing a single flare simultaneously by chance is 0.003\,\%. Similarly, for combinations of four, five, or six instruments the number of flares captured is often a factor of ten or more greater than random. This shows that instrument planners are intentionally co-observing the same flare-productive active regions. % flares are a high priority science target, and that

The analysis in Section~\ref{s:results} suggests that multi-instrument observations are occurring much more frequently than expected by chance. It should be noted that this conclusion is reached by assuming that each instrument is acting independently of all the others. This is a dubious assumption, as the community is aware of the scientific value of multi-instrument observations of solar flares, and many of the people who operate flare-observing instruments have a professional interest in studying flares. It does not take account of known existing community efforts that are designed to promote flare co-observation such as the \textit{Max Millennium Program}, or joint observing programs such as the \textit{Hinode} Operations Program\footnote{\url{www.isas.jaxa.jp/home/solar/guidance/index.html}} that often specifically request that multiple instruments point to the same target. As the number of instruments goes up the deviation towards increased co-observation increases suggesting that the co-observation rate depends not only on which instruments are observing (the $m_{i}$s in Equation \ref{eqn:binomialcombined}) but also on the number of instruments. This suggests that instrument operators are more likely to co-observe if a number of other instruments are already co-observing a target.

Unsurprisingly the instruments with the longest duty cycles and largest fields of view (RHESSI, MEGS-A, XRT) performed the best individually. Reduced field-of-view instruments that require operations planning, while performing within their expected success rates, still only captured on the order of 20\,\% or less of all flares. The possible reasons for this have already been touched upon, but may also be due to other specific factors. For example, until recently, EIS (which observed 6\,--\,8\,\% of flares) only received 15\,\% of \textit{Hinode's} total telemetry since January 2008, thereby limiting its daily duty cycle. This has since been revised up to 23\,--\,43\,\%\footnote{\url{solarb.mssl.ucl.ac.uk/SolarB/hinode_revised_tlm.html}} which should improve its statistics. Similarly, MEGS-B was often only exposed to the Sun for three hours per day to minimize detector degradation, but now responds to a flare trigger when the flux level at 1\,--\,7\,\AA\ as measured by the ESP component of EVE exceeds the GOES M1 level \citep{mill16}. This effectively makes MEGS-B a dedicated flare instrument and it is likely to observe a higher fraction of flares in the future.

\section{Discussion}
\label{s:disc}
This study raises a significant question for the solar-physics community: even though we seem to be co-observing flares at rates larger than those expected by chance, is the number of co-observed flares acceptable or not? This is not an easy question to answer. This study presents some analysis regarding the retrospective behavior of the community regarding co-observation of solar flares, but does not comment on what the solar-physics community wants to do, or should do. It seems obvious that increasing the number of co-observed flares is desirable, but there are three factors for the community to consider in relation to this question.

The first factor is the desire of the community to use limited instrumental resources to study flares compared to other solar features and phenomena. If the community decides that the study of solar flares is relatively more important than the study of other features and phenomena, then their relatively unpredictable occurrence means that a greater fraction of each instrument's observational resources should be devoted to capturing as much data as possible when a flare-productive region appears (\textit{e.g.} in response to a \textit{Max Millennium Major Flare Watch}). Other more ``quiescent'' targets (e.g. coronal holes, filaments, plage regions, etc.) are much more commonplace and can be observed at almost any time.

Having decided that flare research is important, the next factor to consider is our ability to predict when and where a flare will occur. Without a reliable method of predicting when and where a solar flare will occur, we are left with trying to optimize instrumental resources in the face of incomplete information as well as each instruments' operational constraints and competing scientific priorities. The \textit{Max Millennium Program} aims to provide an assessment of the likelihood of a flare in a given region over the following 24 hours. As well as a human assessment of flare likelihood, we suggest that machine-learning techniques be employed as another tool to aid the human flare forecaster. For example, \cite{bobr15} use support vector-machine-methods to determine flare probabilities. Another possible approach is to aggregate results from all existing flare-prediction tools to provide a single, combined measure assessing flare likelihood. It is also fundamentally important that support for the basic science of understanding how and why a solar flare is (or is not) triggered continues.

With a target selected, the final factor to consider is the number of co-observing instruments required to answer the particular science question. The utility of a set of multi-instrument flare observations depends on the science question being asked. All flares do not have to be observed by all instruments all of the time. For example, to understand the dynamic response of the chromosphere to energy deposited by nonthermal electrons, perhaps only RHESSI, EIS, and IRIS data are necessary \citep{bros16}. For flare differential emission measure studies, maybe having simultaneous RHESSI and EVE \citep{casp15} or EIS and XRT observations \citep{odwy14} are desirable. This is the primary function of the IDL widget described in Appendix~\ref{appendixa}: to allow users to quickly and easily search for joint datasets of solar-flare observations in order to answer a specific science question. Assessing the scientific impact of flare co-observations is difficult. It may be possible to measure the scientific impact of flare co-observation through citation analysis (for example, \citealt{desollaprice510, Giles:1998:CAC:276675.276685, kaur}) of flare articles as a function of number of instruments. This type of study is beyond the scope of this article, but it would provide more information to the community in understanding the scientific impact of flare co-observations.

Understanding how a solar flare operates is a fundamental challenge to our understanding of the Sun and the conditions of the heliosphere. In the upcoming Solar Cycle 25, the \textit{Daniel K. Inouye Solar Telescope}, \textit{Solar Orbiter}, and \textit{Parker Solar Probe} will all be operational. These facilities all have limited duty cycles, and different operational constraints. Optimizing the solar-flare science return from these and other instruments relies on continued inter-instrument co-ordination where possible. We suggest that each instrument's observational plans be made available online, ideally in a commonly agreed format. Tools should be developed to read and visualize those plans (the Helioviewer Project clients \textsf{helioviewer.org} and \textsf{JHelioviewer} could be extended to present observation plans) that could take into account solar differential rotation, overplotting it on images from many different instruments. This will enable instrument operators, scientists and other users to understand how and why particular observations are being planned. We suggest that increased planning transparency will inevitably lead to an increased understanding of how an instrument's operations create the \textit{revealed} science priorities of an instrument, as opposed to its \textit{stated} priorities. From this basis a better understanding of how to co-ordinate co-observations between instruments can be generated. Finally, it should be noted that co-observation of \textit{non-flaring} regions is also of considerable scientific value. Co-observations that do not catch a flare are not without merit; much can be learned about the physics of active regions, the chromosphere--corona connection, polarity inversion lines, sunspots, etc, using observations from multiple instruments.

\begin{acks}
R.O. Milligan is grateful for financial support from NASA LWS/SDO Data Analysis grant NNX14AE07G, the Science and Technologies Facilities Council for the award of an Ernest Rutherford Fellowship (ST/N004981/1), and to Kim Tolbert for help with developing the \textsf{SSWIDL} widget. J. Ireland acknowledges gratefully the support of the Heliophysics Data Environment Enhancement program and the Solar Data Analysis Center. Both authors are also very grateful to the anonymous referee who provided critical and constructive feedback that greatly improved the quality and scope of this article. Error-bar estimates on the expected number of flares were calculated using \textsf{SciPy} \citep{scipy}.
\end{acks}

\section*{Disclosure of Potential Conflicts of Interest}
The authors declare that they have no conflicts of interest.

\appendix

\section{Solar Flare Finder Widget}
\label{appendixa}

\begin{figure}[!t]
\begin{center}
\includegraphics[width=\textwidth]{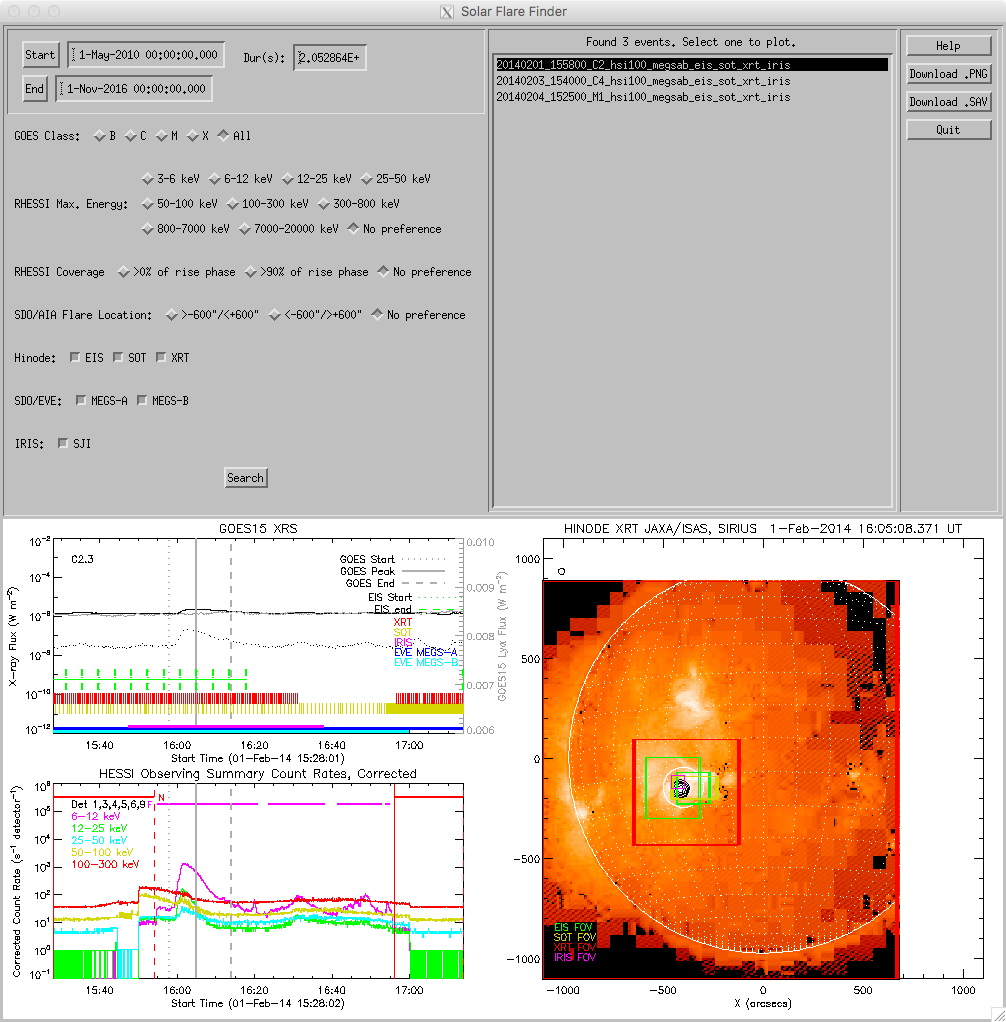}
\caption{Screenshot of the Solar Flare Finder widget in \textsf{SSWIDL}. The sample flare shown is an C2.3 flare that was observed simultaneously by all instruments on 4 February 2014.}
\label{sff_idl}
\end{center}
\end{figure}

The list of solar flares used for the analysis in this article were used as the source data for the \textsf{Solar Flare Finder}, a widget that was developed in tandem with this study that allows users to search for flares observed by a chosen set of instruments. The Solar Flare Finder widget is available now via \textsf{SolarSoftWare/IDL} (\citealt{free98}; \textsf{idl> solar\_flare\_finder}). A screenshot is shown in Figure~\ref{sff_idl}. The widget searches a pre-generated lookup table\footnote{\url{hesperia.gsfc.nasa.gov/sff/ssw_sff_list.txt}} (the same lookup table used to generate the \textsf{UpSet R} plots in Figure~\ref{f:upset}, only with B-class flares included and which is continuously being updated) to return SSW Latest Events simultaneously observed by selected instruments. The widget allows the user to search by GOES class (B, C, M, X), flare location (disk; $>-600''\,--\,<+600''$ or limb; $<-600''\,--\,>+600''$), percentage of the rise phase covered by RHESSI ($>$0\,\% or $>$90\,\%), and by the maximum energy recorded by RHESSI. The widget returns a list of flares conforming to the users specifications (if any), allowing the user to click on a desired event to bring up a plot similar to that shown in Figure~\ref{sff_plot} that displays the metadata from all available datasets. These plots, and the associated metadata (in the form of an IDL .sav file), are downloadable via the widget, and are hosted at \url{http://hesperia.gsfc.nasa.gov/sff/}. Note that no guarantees are made regarding the quality of the data itself. Perhaps the EIS or IRIS slits may not have been precisely aligned with the flare ribbons during the impulsive phase, or the solar background may have been sufficiently high that, say, $<$C5 flares do not show up in EVE data. However, this tool aims to greatly narrow the search for specific events that match a user's request in order to answer particular science questions. 

\section{Expected Versus Measured Number of Flares}
\label{appendixb}

The five panels in Figure~\ref{fig:pred_succ} compare the expected number of flares [$\enf$: blue triangles] of a given combination of instruments observing a flare (based on individual measured success rates, $\measuredsuccessrate$, from Table~\ref{tab:estimated_flares} and Equation~\ref{eqn:binomialcombined}) with the actual measured values [$\mnf$, red diamonds] during the 11-month period for which all seven of the instruments were operating. The ratio of these values [$\mnf/\enf$: solid black circles] indicates how successful each given combination has performed. A value greater than unity indicates that a given combination has performed better than random. For almost all combinations of instruments (particularly higher $r$ values; $>$4), $\mnf>>\enf$. Furthermore, if we consider the probability of all seven instruments targeting the same flare independently, then the product of the individual expected values gives us a probability of 0.003\,\%. Therefore the measured value of three flares (0.3\,\% of 934) is actually considerably better than what one might expect for such a fortuitous combination of observations.

\begin{figure}[!h]
\begin{center}
\subfloat{\includegraphics[width=0.97\textwidth]{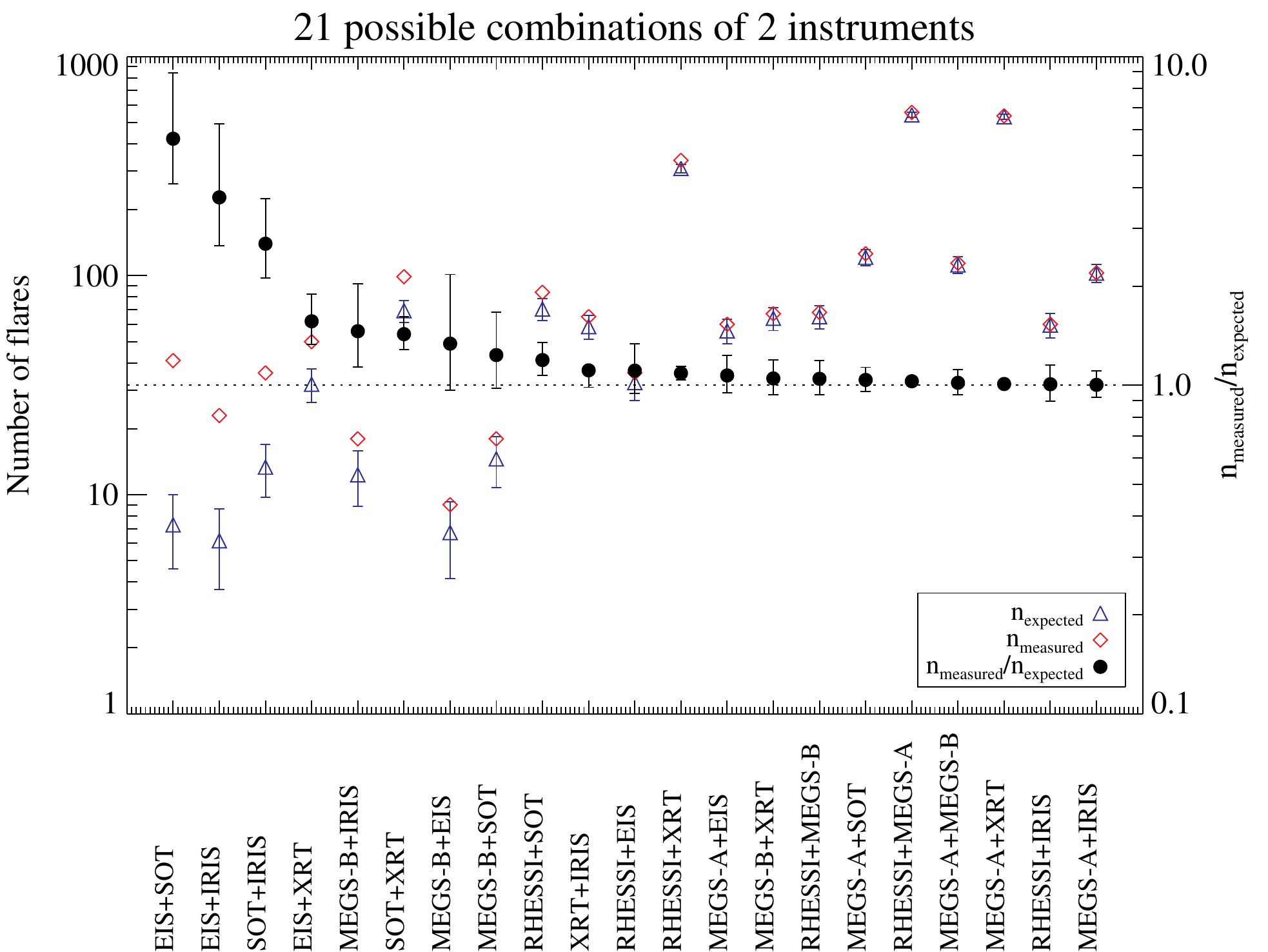} } \,
%\subfloat{\includegraphics[width=0.97\textwidth]{predicted_inter_mission_success_3insts.eps} } \,
\caption{The measured (red diamond) and expected (blue triangle) probabilities of any given combination of (2, 3, 4, 5, or 6 out of 7) instruments successfully observing a solar flare. The solid-black circles denote the ratio of measured to expected rates.}
\label{fig:pred_succ}
\end{center}
\end{figure}

\begin{figure}[!t]
\begin{center}
\ContinuedFloat
\subfloat{\includegraphics[width=0.97\textwidth]{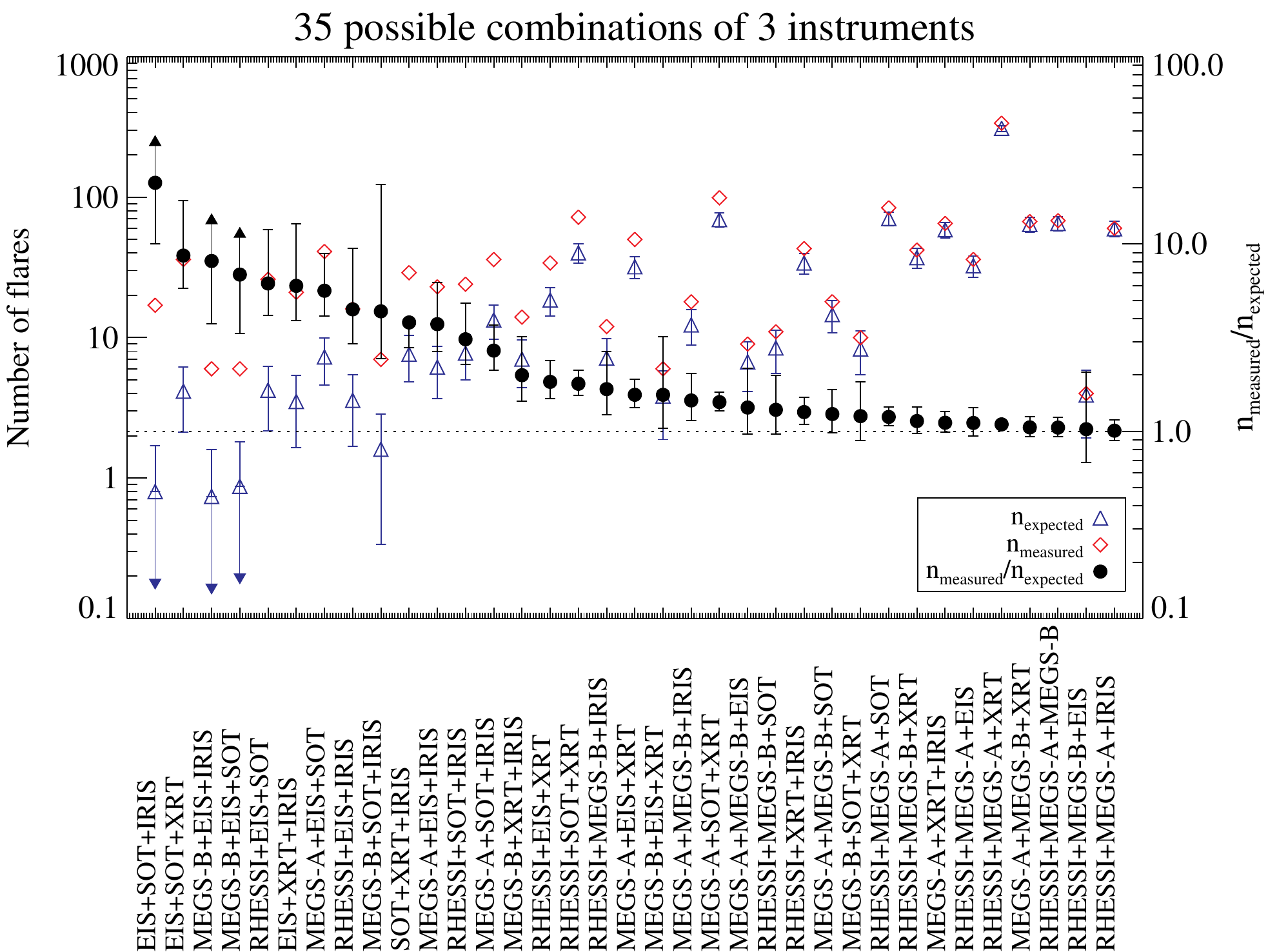} } \,
\subfloat{\includegraphics[width=0.97\textwidth]{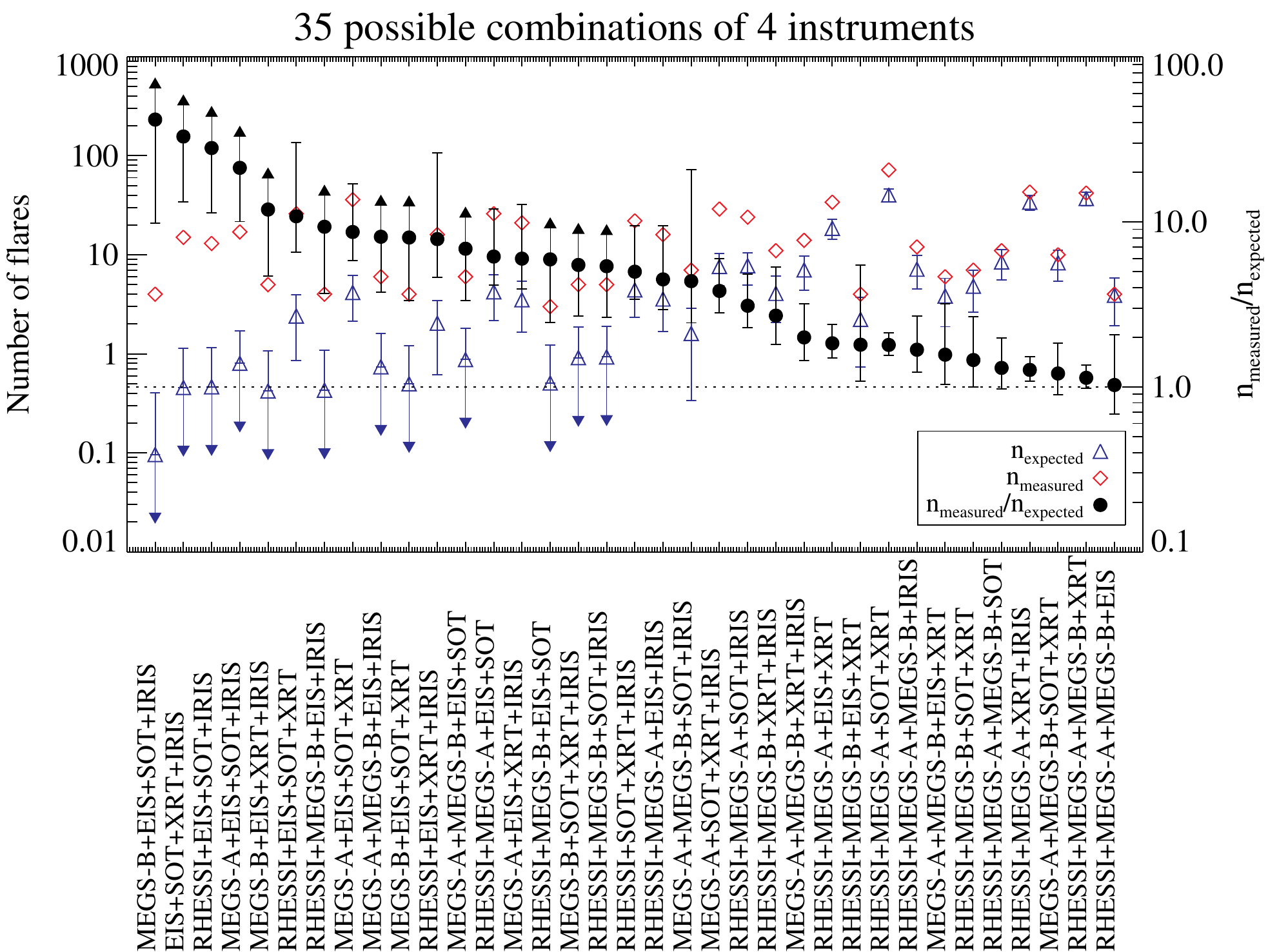} } \,
\caption{The measured (red diamond) and expected (blue triangle) probabilities of any given combination of (2, 3, 4, 5, or 6 out of 7) instruments successfully observing a solar flare. The solid-black circles denote the ratio of measured to expected rates.}
\end{center}
\end{figure}
\begin{figure}[!t]
\begin{center}
\ContinuedFloat
\subfloat{\includegraphics[width=0.97\textwidth]{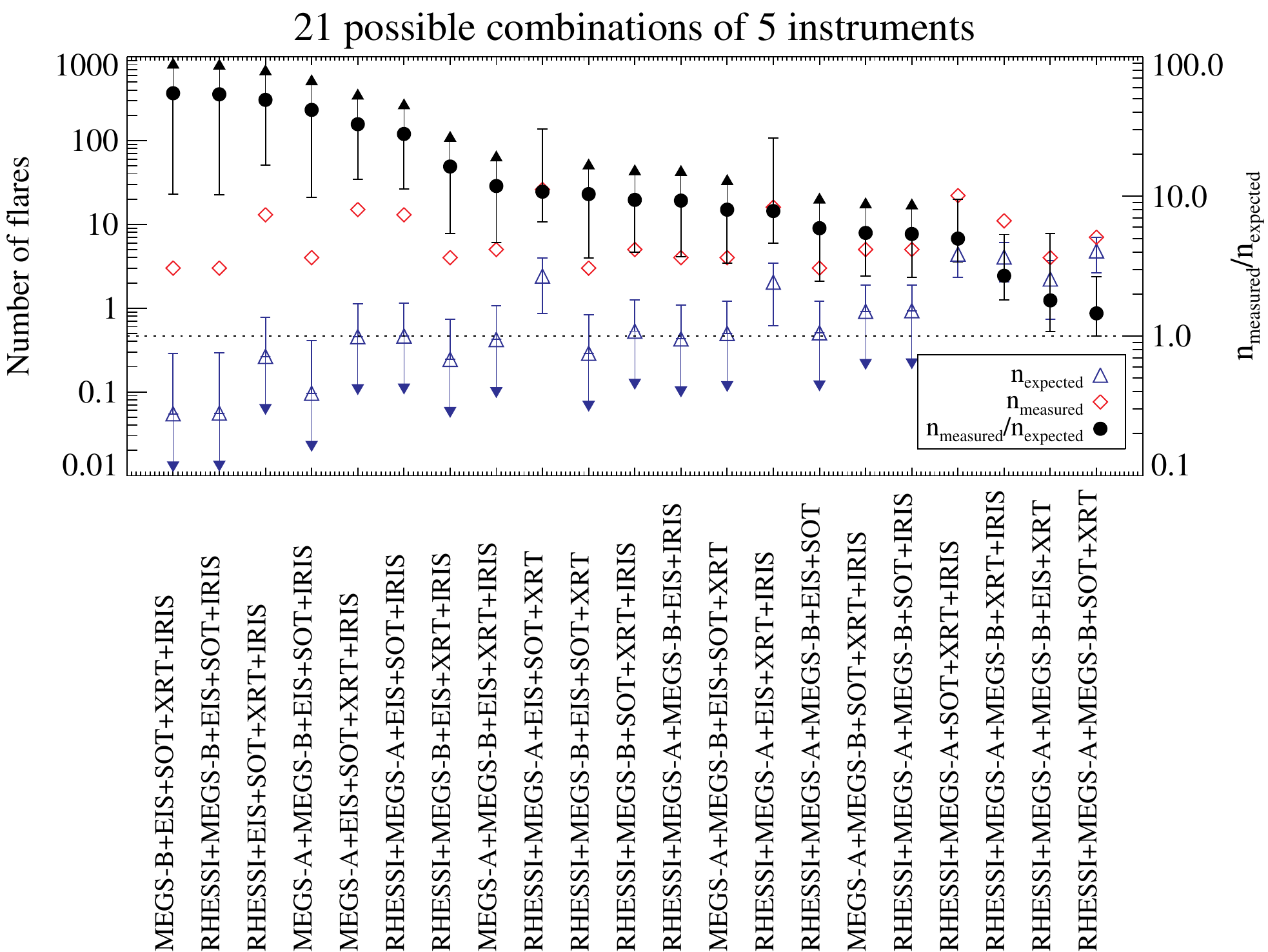} } \,
\subfloat{\includegraphics[width=0.97\textwidth]{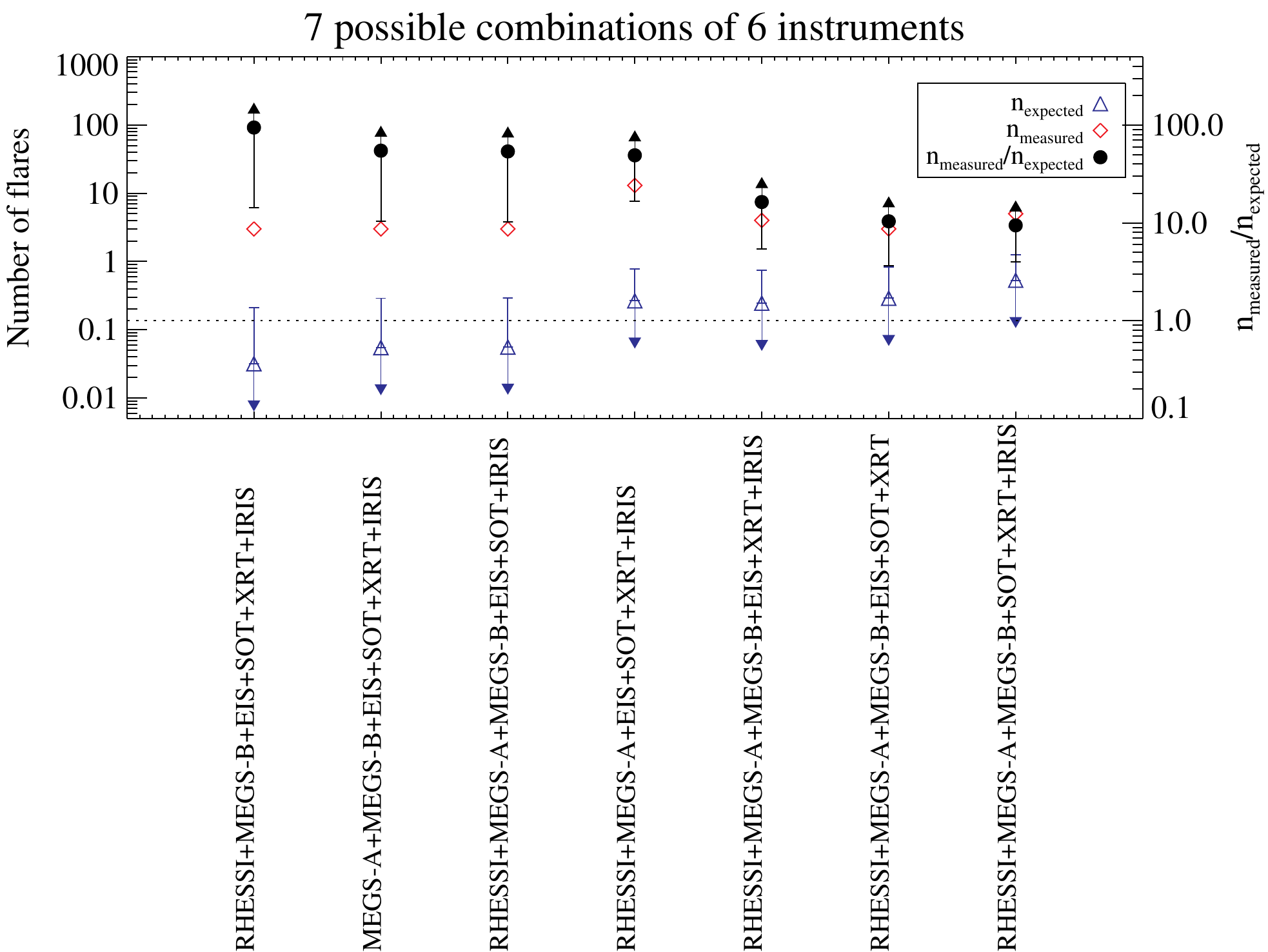} } \,
\caption{The measured (red diamond) and expected (blue triangle) probabilities of any given combination of (2, 3, 4, 5, or 6 out of 7) instruments successfully observing a solar flare. The solid-black circles denote the ratio of measured to expected rates.}
\end{center}
\end{figure}

\newpage
\clearpage

% Using BibTeX
% \begin{thebibliography}{}
%\bibliographystyle{spr-mp-sola}
%\bibliographystyle{spr-mp-sola-cnd} %% Alternative style: no title, no concluding page
%\bibliography{ms}  
% \end{thebibliography}{}

\

\end{article} 
\end{document}